\documentclass[prl,superscriptaddress,twocolumn,amsmath,amssymb]{revtex4}
\usepackage{psfrag}
\usepackage{graphicx}
\usepackage{bm}
\usepackage{subfigure}

\begin{document}

\title{Stability of half-quantum vortices in $p_x+ip_y$
superconductors}

\author{Suk Bum Chung}
\affiliation{Department of Physics, University of Illinois at
Urbana-Champaign, Urbana, Illinois 61801, USA}
\author{Hendrik Bluhm}
\affiliation{Department of Physics, Stanford University, Stanford, California
94305, USA}
\author{Eun-Ah Kim}
\affiliation{Department of Physics, Stanford University, Stanford, California 94305, USA}

\date{\today}

\begin{abstract}
We consider the stability conditions for half-quantum vortices in quasi two dimensional $p_x\!+\!ip_y$ superconductor (such as ${\rm Sr_2 RuO_4}$ is believed to be). The predicted exotic nature of these excitations has recently attracted much attention but they have not been observed yet.
We emphasize that an isolated half quantum vortex has a divergent energy cost in the bulk due to its  {\it unscreened} spin current, which requires two half-quantum vortices with opposite spin winding to pair.
We show that the stability of such a pair is enhanced when the
 ratio of spin superfluid density to superfluid density $\rho_{{\rm sp}}/\rho_{\rm s}$ is small.
 We propose  using various mesoscopic geometries to stabilize and observe these exotic excitations.

\end{abstract}

\maketitle

The possibility of half-quantum vortices ($1/2$-qv's)
in $p_x+ip_y$ superconductors (SC's) has recently added a new `spin' to the interest in such exotic SCs.
The prediction that such vortices in (quasi-) two dimensional (2D) superfluids will have  Majorana fermion zero modes bound at vortex cores, which render vortex statistics non-Abelian\cite{ivanov,read-green,sarma-tewari,tewari07}, drives the excitement.
A realization of excitations with such exotic statistics is
of interest on its own right. Moreover, the possibility of exploiting non-abelian statistics
for topological quantum computation\cite{kitaev} adds technological interest as well.

$1/2$-qv's are topologically allowed in triplet superfluids due to their spin degrees of freedom.
They were first sought after in thin films of $^3$He-A\cite{salomaa}: the best known example of $p_x\!+\!ip_y$ superfluid\cite{leggett_rmp}.
However,  achieving a sufficiently thin parallel plate
geometry is challenging and NMR experiments on $^3$He-A
thin films failed to detect
$1/2$-qv's\cite{hakonen89}.
${\rm Sr_2 RuO_4}$
has recently emerged as a candidate material for a
quasi-2D ($ab$ plane) $p_x\!+\!ip_y$ spin-triplet SC~\cite{rice,ishida_triplet,TRSBex,mackenzieRMP}, offering an alternate system to look for $1/2$-qv's.
Refs~\cite{sarma-tewari,tewari07} proposed
experiments for
probing and exploiting the exotic nature of the Majorana fermion core states of $1/2$-qv's.
Nevertheless, $1/2$-qv's have not been observed in  ${\rm Sr_2 RuO_4}$~\cite{vortex_lattice,hasselbach,per-kam}.
For $1/2$-qv's to be realized, it is crucial to carefully consider their energetics and schemes for stabilization.

In this letter, we investigate the stability of the $1/2$-qv's and discuss routes for their realization using mesoscopic samples.
Vortex energetics in the context of ${\rm Sr_2 RuO_4}$ were first considered by Kee {\it et al.}\cite{kee_sr}  for 
$1/2$-qv's in the $bc$ plane 
bound to a texture in the {\bf d} vector (Eq.~\eqref{EQ:dVec}).
With the interest for the exotic statistics (which is only possible in 2D) in our minds, we focus on the $ab$ plane vortices.
Das Sarma {\it et al.} noted \cite{sarma-tewari} that an
out-of-plane field of the order of the spin-orbit decoupling field\cite{murakawa} can reduce the energy cost of $1/2$-qv's in ${ab}$ plane\cite{sarma-tewari}. 
They further argued that the smaller vorticity of 1/2 qv's
may energetically favor them over full qv's.
However, as it was remarked in Ref.\cite{sigrist_rmp},
an isolated $1/2$-qv costs energy that is divergent in the system size due to its {\it unscreened} spin current. Indeed, fractional vortices that are allowed in multi-component SC generally tend to be energetically unfavorable \cite{babaev}. Hence free isolated $1/2$-qv's cannot exist in the bulk.
There are two ways to cut off this divergence:
1) two $1/2$-qv's with opposite spin winding forming a pair in a bulk sample, 2) a $1/2$-qv entering a mesoscopic sample. Through explicit energetics calculations, we find that $1/2$-qv's can be stabilized when the ratio $\rho_{{\rm sp}}/\rho_{\rm s}$  is sufficiently small.

{\it Stability Analysis} -- The order parameter of a triplet SC takes a
matrix form in the spin space~\cite{mackenzieRMP,sigrist_rmp}:
\begin{equation}
\hat{\Delta} ({\bf k}) \!=\! \left [
\begin{array}{cc}
  \Delta_{\uparrow\uparrow}({\bf k}) & \Delta_{\uparrow\downarrow}({\bf k}) \\
  \Delta_{\downarrow\uparrow}({\bf k}) & \Delta_{\downarrow\downarrow}({\bf k})\\
\end{array}
\right ]
\!\equiv\! \left[
  \begin{array}{cc}
  -d_x + id_y & d_z \\
  d_z & d_x + id_y \\
\end{array}
\right ].
\label{EQ:dVec}
\end{equation}
The triplet pairing requires $\Delta_{\uparrow\downarrow}\!=\!
\Delta_{\downarrow\uparrow}$
and the ${\bf k}$ dependent vector ${\bf d}({\bf k})$ was introduced to parameterize the gap function.
For each ${\bf k}$, the unit vector ${\bf\hat d}({\bf k})$ represents the
symmetry direction with respect to the rotation of Cooper pair
spin.
In a $p_x+ip_y$ SC, the ${\bf k}$ dependence of ${\bf d}({\bf
k})$ is determined by the pairs having
finite angular momentum projection $m=1$ directed along ${\bf \hat l}$.
Hence
\begin{equation}
{\bf d} ({\bf k}) = \Delta_0 {\bf {\hat d}} \exp (i\varphi_{{\bf{\hat k}}{\bf{\hat l}}}) |{\bf{\hat k}}\times{\bf{\hat l}}|,
\label{EQ:OPp+ip}
\end{equation}
where $\varphi_{{\bf{\hat k}}{\bf{\hat l}}}$ is the azimuthal angle ${\bf{\hat k}}$ makes around ${\bf{\hat l}}$.
For a (quasi-)2D system with ${\bf {\hat l}}=\pm{\hat z}$ 
, Eq.~\eqref{EQ:OPp+ip}  simplifies:
\begin{equation}
{\bf d} ({\bf k}) = \Delta_0 {\bf {\hat d}} \exp (i\varphi_{{\bf{\hat k}}}),
\label{EQ:OPp+ip2}
\end{equation}
for a single domain,  where $\varphi_{{\bf{\hat k}}}$ is the azimuthal angle ${\bf{\hat k}}$ makes around the $c$-axis.
Notice two
independent continuous symmetries associated with ${\bf d} ({\bf k})$ :
the spin rotation symmetry $SO_2$ around  ${\bf
{\hat d}}$, and the $U(1)$ symmetry combining  a gauge
transformation and an orbital rotation around ${\bf {\hat
l}}$~\cite{vollhardt_book}.  For fixed ${\bf{\hat
l}}$, we denote the `orbital phase' for  this combined
$U(1)$ symmetry by $\chi$.

The spin degree of freedom represented by the ${\bf d}$-vector
allows for the existence of half-quantum vortices.
In a singlet (single component) superconductor, $1/2$-qv's are forbidden in order for the order parameter to be single valued.
However in the $p_x+ip_y$ SC, each component of the order parameter matrix can remain single valued by simultaneously rotating the ${\bf d}$-vector by $\pi$ while the orbital phase $\chi$ winds by $\pi$.
Since only $\chi$ couples to the magnetic vector potential,
this type of topological defect only carries a flux of $hc/4e$,
i.e. half a superconducting flux quantum $\Phi_0 = hc/2e$.
From Eq.~(\ref{EQ:dVec}), one can
intuitively characterize a $1/2$-qv as a vortex with a single unit of vorticity for one of the spin components and zero vorticity for the other~\cite{leggett-yip}.

In the absence of an external magnetic field, the spin-orbit
(dipole) coupling favors alignment of ${\bf {\hat d}}$ and ${\bf {\hat
l}}$,
making $1/2$-qv's energetically costly~\cite{vollhardt_book,salomaa}.
However, Knight-shift measurement on ${\rm Sr_2 RuO_4}$ by Murakawa {\it et
al.}~\cite{murakawa} suggests that a sufficiently large magnetic field
$H > $200 G ($<H_{c2}\sim 750$ G \cite{mackenzieRMP}) along the $c$-axis
may neutralize the dipolar interaction by fixing the
${\bf d}$-vector orientation to the $ab$-plane (or ${\rm RuO_2}$ plane).
While this opens the possibility of $1/2$-qv's,
the associated ${\bf d}$-vector bending introduces `hydrodynamic' spin terms in the
gradient  free energy.
In the following, we will determine their effect on the stability of
half-quantum vortices.

The Ginzburg-Landau
(GL) gradient free energy in its most general form in the London
limit (i.e., the superfluid density is taken to be constant outside vortex cores~\cite{vollhardt_book}):
\begin{align}
&f_{{\rm grad}} = \frac{1}{2}[ \rho_{\rm s} v_s^2 - (\rho_{\rm s} -
\rho_{\rm s}^\|) ({\bf {\hat l}} \cdot {\bf v_s})^2 ]\nonumber\\
&\;+\! \frac{1}{2}
\left(\frac{\hbar}{2m}\right)^2\sum_i[\rho_{\rm sp} (\nabla {\hat
d}_i)^2 - (\rho_{\rm sp} - \rho_{\rm sp}^\|) ({\bf {\hat l}}
\cdot \nabla {\hat d}_i)^2]\nonumber\\ &\;+\! K_{ij}^{mn}
\partial_i {\hat l}_m \partial_j {\hat l}_n \!+\! C_{ij} (v_{\rm s})_i
(\nabla \times {\hat l})_j \!+\!\frac{1}{8\pi}(\nabla\!
\times\!\textbf{A})^2,
\label{EQ:gradGen}
\end{align}
where ${\bf v_s}$ is the superflow velocity, $m$ is the fermion mass,
and $\rho_{\rm s}$, $\rho_{\rm s}^\|$ and $\rho_{\rm sp}$, $\rho_{\rm
sp}^\|$ are components of the superfluid and spin fluid density
matrix, respectively.  Note that these are rank-two matrices, since
there is one symmetry direction, ${\bf {\hat l}}$, for the orbital degree
of freedom.  For a 2D single-domain, the fact that ${\bf {\hat l}}$ is fixed leads to great simplification.  The superflow velocity ${\bf
v_s}$ is then $(\hbar/2m)(\nabla \chi - 2e{\bf A}/\hbar c)$, as in a
conventional $s$-wave SC.  For vortex lines along the $c$-axis,
$\rho_{{\rm s}\|}$ and $\rho_{{\rm sp}\|}$ are projected out of the
problem due to the absence of any variation along the ${\bf {\hat l}}$
direction ($c$-axis), and the superfluid and spin current densities can
be regarded as scalars.  For  ${\bf {\hat d}} = (\cos \alpha, \sin \alpha,0)$ in  the $ab$-plane the
free energy becomes
\begin{equation}
f^{{\rm 2D}}_{{\rm grad}}\! =\! \frac{1}{2}\! \left(\frac{\hbar}{2m}\right)^2\!\!
\left[\rho_{\rm s}\! \left(\!\nabla\!\!_\bot\! \chi \!-\! \frac{2e}{\hbar
c}\textbf{A}\!\right)^2\!\! +\! \rho_{{\rm sp}} \left(\nabla\!\!_\bot\! \alpha\right)^2\!\right]
\!+\! \frac{1}{8\pi}(\nabla\! \times\! \textbf{A})^2,
\label{EQ:gradientFree1}
\end{equation}
with the last term accounting for screening of the supercurrent, which
is absent for the case of $^3$He-A thin film~\cite{salomaa,kee_he3}.

\begin{figure}[t]
\psfrag{x}{$r_{12}/\lambda$}
\psfrag{y}{$E_{{\rm pair}}^{{\rm half}}(r_{12})\!-\!E_{{\rm full}}$}
\subfigure{
\includegraphics[width=.14\textwidth]{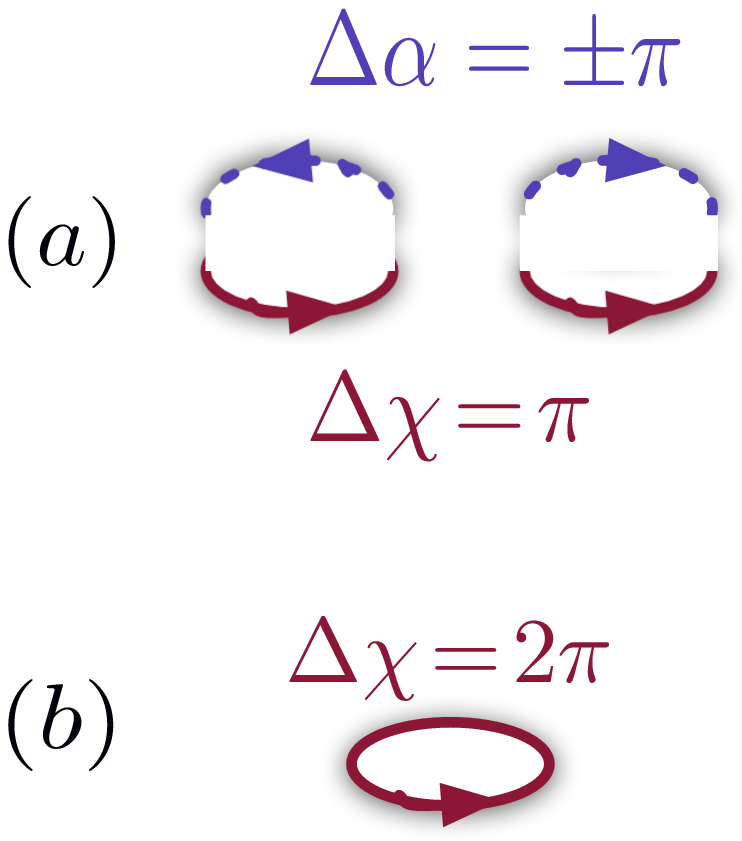}
}
\subfigure{
\includegraphics[width=.22\textwidth]{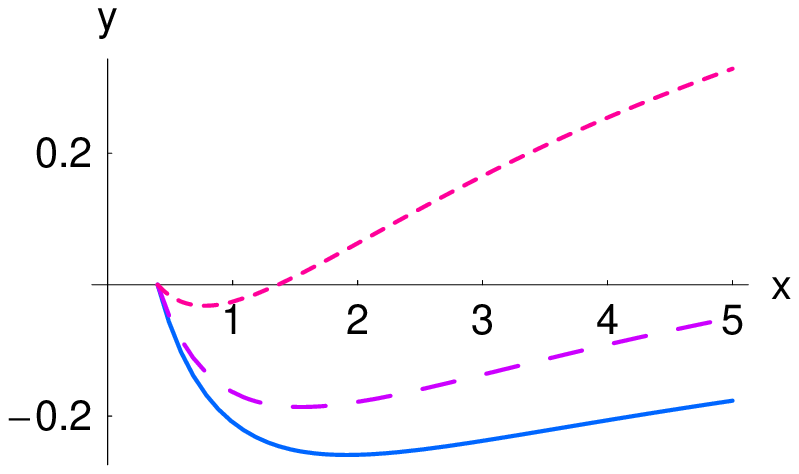}
}
\caption{The energy difference (see the text) as a function of pair
size, using $\xi = 0.4\lambda$ for ${\rm Sr_2 RuO_4}$ for different
values of $\rho_{{\rm sp}}/\rho_{\rm s}$: $\rho_{{\rm sp}}/\rho_{\rm
s}= 0.3$ (the blue solid curve), $\rho_{{\rm sp}}/\rho_{\rm s}= 0.4$
(the purple dashed curve), and $\rho_{{\rm sp}}/\rho_{\rm s}= 0.7$
(the red dotted curve).  The energy per unit length is in units of
$\Phi_0^2 / 16\pi^2 \lambda^2$.  }
\label{Fig:pair}
\end{figure}

Unlike a full qv, 
a single $1/2$-qv costs a spin current energy
 that diverges logarithmically\cite{sigrist_rmp}
\begin{equation}
 \epsilon_{{\rm sp}}=\frac{\pi}{4} \left(\frac{\hbar}{2m}\right)^2 \rho_{{\rm sp}} \ln \left(\frac{R}{\xi}\right)
\label{EQ:dLine}
\end{equation}
per unit length, where $\xi$ is the core
radius and $R$ the lateral sample dimension.
This divergence is due to the {\it absence of screening}, for
 the ${\bf \hat{d}}$-winding
(the $\nabla \alpha$ term in Eq.~(\ref{EQ:gradientFree1})) does not
couple to the electromagnetic field.

A pair of $1/2$-qv's with
opposite sense
windings in ${\bf\hat{d}}$ 
would be free of such divergent energy cost.
The energy
per unit length of a  pair of $1/2$-qv's separated by $r_{12}$ is
\begin{equation}
E_{{\rm pair}}^{{\rm half}}(r_{12})\! =\!\frac{1}{2} \frac{\Phi_0^2}{16 \pi^2
\lambda^2} \!\left[\!\ln \left(\!\frac{\lambda}{\xi}\!\right)\! +\!
K_0\! \left(\!\frac{r_{12}}{\lambda}\!\right)\! +\! \frac{\rho_{{\rm
sp}}}{\rho_{\rm s}} \ln \left(\!\frac{r_{12}}{\xi}\!\right)\!\right],
\label{EQ:half}
\end{equation}
where $K_0$ is a modified Bessel function and $\lambda\! =\! \frac{mc}{2e {\sqrt {\pi\rho_s}}}$ is the London
penetration depth.
The first term of Eq.~(\ref{EQ:half}) is the {\it
screened} self-energy of superflow.  The second term
accounts for the magnetic interaction between the
vortices, and the third term, in which the logarithmic divergence
of Eq.~(\ref{EQ:dLine}) is canceled out by the interaction term,
is purely due to spin flow.
However, such a pair
is topologically equivalent to a
single full qv (see Fig.~\ref{Fig:pair}(a-b)) with the self-energy
\begin{equation}
E^{{\rm full}} \!=\!\pi \left(\frac{\hbar}{2m}\right)^2\!\! \rho_{{\rm s}} \ln \left(\frac{\lambda}{\xi}\right)
\!=\! \frac{\Phi_0^2}{16 \pi^2 \lambda^2} \ln \left(\frac{\lambda}{\xi}\right).
\label{EQ:full}
\end{equation}
Hence the most relevant question, regarding the stability of $1/2$-qv's in the
bulk of a SC is whether a pair of $1/2$-qv's can be generated by
the decay of a full qv.
 (Note that
Eqs.~(\ref{EQ:half}-\ref{EQ:full}) are approximately equal at
$r_{12}\!=\! \xi$.)\footnote{Eq.\eqref{EQ:half}  suggests an entropy driven transition of
Berezinskii-Kosterlitz-Thouless type at $T_{{\rm KT}}$ below $T_c$.
The onset temperature is $k_B T_{{\rm KT}}\!=\!\frac{1}{8}
\frac{T_c - T_{{\rm KT}}}{T_c}
\frac{\rho_{{\rm sp}}}{\rho_{\rm s}} d
\left(\frac{\Phi_0}{4\pi\lambda}\right)^2$,
where $d$ is the sample thickness. A related discussion can be found in E.\ Babaev, Phys.\ Rev.\ Lett.\ {\bf 94}, 137001 (2005).
However, experimental values of $\lambda$ and $T_c$ suggests $T_{{\rm KT}} \approx T_c$.}

We find that a $1/2$-qv pair in bulk ${\rm Sr_2 RuO_4}$ will only be stable if
$\rho_{{\rm sp}}/\rho_{\rm s}$ satisfies certain conditions.
 In Fig.~\ref{Fig:pair}, we plot
$E_{{\rm pair}}^{{\rm half}}(r_{12})-E^{{\rm full}}$ for $r_{12}>\xi$.  We used $\kappa \equiv
\lambda/\xi = 2.5$ based on  $\lambda_{ab}
\approx 152$ nm~\cite{vortex_lattice,mackenzieRMP} and the
estimated value of the Ginzburg-Landau coherence length $\xi_{ab} \sim
66$ nm~\cite{mackenzieRMP}.  For all three values of $\rho_{{\rm
sp}}/\rho_{\rm s}$ shown, a finite equilibrium pair size
$r_{{\rm equil}}>\xi$ exists.
However, we note that
while the London approximation predicts a minimum of the energy difference $E_{{\rm pair}}^{{\rm
half}}(r_{12})-E^{{\rm full}}$ at a finite value $r_{12}\!=\!r_{\rm
equil}$ for $\rho_{{\rm sp}}/\rho_{\rm s} < 1$,
the London limit is only valid when
 $\rho_{{\rm sp}}/\rho_{\rm s}$ is small enough to give
$r_{{\rm equil}}>\xi$.

We now discuss the ratio $\rho_{{\rm sp}}/\rho_{\rm s}$ between
 the neutral spin superfluid density
$\rho_{{\rm sp}}$
and the charged mass superfluid density $\rho_{{\rm s}}$  in Eq.~(\ref{EQ:gradientFree1}) .
In the case of $^3$He, this ratio has been derived theoretically in
terms of Landau
parameters~\cite{leggett_rmp,vollhardt_book,leggett_interaction,kee_he3}.
Using sum rules for the longitudinal current-current correlation, Leggett
showed that $\rho_{{\rm sp}}<\rho_{\rm s}$~\cite{leggett_interaction}.
This follows from the fact that the mass current is
conserved even in the presence of interactions due to Galilean
invariance, while the spin current is not, since a corresponding
symmetry is absent.  In the case of Sr$_2$RuO$_4$, the lattice breaks the
Galilean invariance.
However, since  all interactions that scatter the mass current also must
scatter spin current, but not vice versa,  one still expects
$\rho_{{\rm sp}} < \rho_{\rm s}$~\cite{leggett_interaction}.
While the actual value of $\rho_{{\rm sp}}/\rho_{\rm s}$ is unknown for
${\rm Sr_2 RuO_4}$, it has recently been
measured to be $\sim 0.3$ near $T=0$ in $^3$He-A~\cite{kee_he3}.
Since many Fermi liquid properties of ${\rm Sr_2 RuO_4}$ are similar to that of
$^3$He~\cite{mackenzieRMP}, $\rho_{{\rm
sp}}/\rho_{\rm s}$ for ${\rm Sr_2 RuO_4}$ is also not anticipated to be much larger
than $\sim 0.4$. If so, the $1/2$-qv pair would have a robust
energy minimum (see Fig.~\ref{Fig:pair}).

Although the above analysis implies that a pair of half quantum vortices can be stable against {\it combining} into a single full qv,
 the vortex  core energy  can prevent a single full qv from {\it decaying} into a pair of half quantum vortices.
Since the core interactions may be substantial for
$r_{12} \lesssim \xi$, it is not
guaranteed that the finite equilibrium separation in our analysis corresponds to a
global energy minimum. When
a finite separation
$r_{{\rm equil}}>0$ configuration only represents a metastable state, its formation
depends on the history of the sample. In this case, a full vortex would not
decay into two $1/2$-qv's.
If $\rho_{{\rm sp}}/\rho_{{\rm s}}$ is so large that $r_{\rm equil}
\lesssim \xi$,  even a local minimum might be absent.
However if $\rho_{{\rm sp}}/\rho_{{\rm s}}\sim 0.4$, as in the case of $^3$He, this is rather unlikely.

{\it Mesoscopic geometries} --
Given that the experimentally unknown core interaction may be significant, and neutron scattering~\cite{vortex_lattice} and vortex imaging~\cite{per-kam,hasselbach}  found no
evidence for
vortex dissociation in macroscopic samples,
we propose utilizing mesoscopic
geometries.
 A sample size of order $\lambda$ would promote the chance
of a $1/2$-qv entering by cutting off the divergent spin current
energy and reducing the effect of supercurrent screening.
Here we discuss a thin slab and a cylinder as two examples.

\begin{figure}[b!]
\subfigure[]{
\includegraphics[width=.13\textwidth]{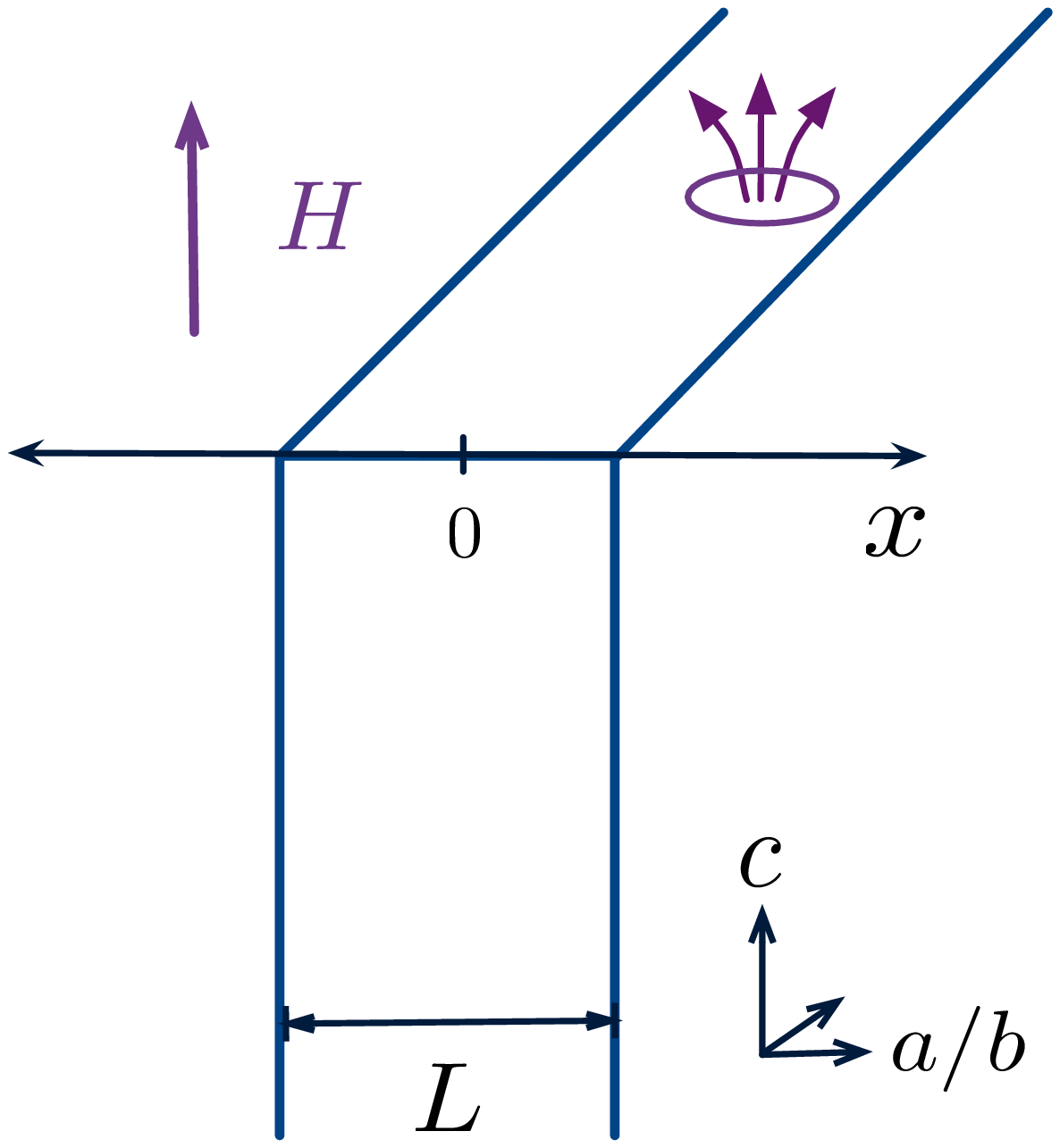}
}
\subfigure[]{
\psfrag{x}{$x/\lambda$}
\psfrag{y}{$G[\frac{\Phi_0^2}{(4\pi\lambda)^2}]$}
\includegraphics[width=.3\textwidth]{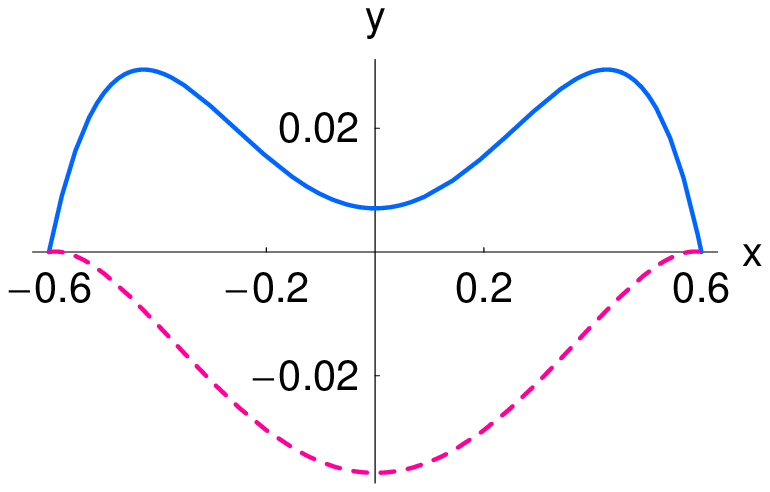}
}
\caption{ (a)A slab of thickness $L\!=\!
2\lambda\!=\!0.304\mu$m in an external field $H$.  (b) $G(x)$ (see
Eq.~(9)) in units of $\Phi_0^2/(4\pi\lambda)^2$ for a full-qv (blue
solid curve) and for a $1/2$-qv (red dashed curve) at $H\!=\!212
G\sim3.0\Phi_0/4\pi\lambda^2$, for $\xi = 0.4\lambda$ and $\rho_{{\rm
sp}}/\rho_{{\rm s}} = 0.4$.}
\label{Fig:slab}
\end{figure}

We find that a slab of thickness $L\!=\!2\lambda$ (see
Fig.~\ref{Fig:slab}) will only permit the entry of $1/2$-qv if $\rho_{{\rm sp}}/\rho_{\rm s} <
(2\xi/\lambda)K_1(2\xi/\lambda)$.
Consider the Gibbs free energy for a vortex at position $x$ in the slab (see Fig.~\ref{Fig:slab}),
with  winding numbers $n_{{\rm s}}$ and $n_{{\rm sp}}$ in $\chi$ and $\alpha$, respectively, and a
uniform applied field $H$:
($n_{{\rm s}}\!=\!n_{{\rm sp}}\!=\!1/2$ and $n_{{\rm s}}\!=\!1$,
$n_{{\rm sp}}\!=\!0$ for a $1/2$-qv
and a full qv, respectively)
\begin{eqnarray}
G(x;\, n_{{\rm s}},n_{{\rm sp}}) \!&=&\! \varepsilon_{{\rm m}}(x;\, n_{{\rm s}})\nonumber\\
\!&+&\! \sum_{\delta=\mp}\! \sum_{j\!=\!1}^\infty\! (-1)^{j-1}\varepsilon_{{\rm bc}}\!(|x\!-\!x_j^\delta|;\, n_{{\rm s}}\!,n_{{\rm sp}}).
\label{eq:G}
\end{eqnarray}
Here, $\varepsilon_{{\rm m}}(x;\, n_{{\rm s}})$ is the energy (per unit sample thickness) due to the interaction
with the Meissner current:
\begin{equation}
\frac{\varepsilon_{{\rm m}}(x;\, n_{{\rm s}})}{\Phi_0^2/(4\pi\lambda)^2} = n_{\rm s} \frac{H}{\Phi_0/4\pi\lambda^2}
\frac{\cosh(x/\lambda)-\cosh[(L-2\xi)/2\lambda]}{\cosh(L/2\lambda)}.
\label{EQ:MeissnerForce}
\end{equation}
The second line of Eq.~\eqref{eq:G} accounts for the boundary
condition of vanishing current normal to each surface.  We solve the
boundary condition using an infinite set of image vortices at positions
$x_j^\mp\!\equiv\! (\!-\!1)^j (x'\! \mp\! jL)\!$ for $j\!=\!1,\ldots,\infty$ with vorticities
$(-1)^j n_{\rm s, sp}$.
Between a vortex with winding number $(n_{{\rm s}},n_{{\rm sp}})$ in the slab and each of its image vortices at a distance $r$ from it,
there is an interaction energy (per unit sample thickness):
\begin{equation}
\frac{\varepsilon_{{\rm bc}}(r;\, n_{{\rm s}},n_{{\rm sp}})}{\Phi_0^2/(4\pi\lambda)^2}
\!=\!-\!n_{\rm s}^2\! \left[\!K_0\!  \left(\frac{r}{\lambda}\right)\!-\!K_0\!  \left(\frac{2\xi}{\lambda}\right)\!\right]
\!+\! n_{{\rm sp}}^2\!\frac{\rho_{{\rm
	sp}}}{\rho_{\rm s}}\!\ln\!\left(\frac{r}{2\xi}\right)\ .
\label{EQ:imageForce}
\end{equation}

We plot $G(x;\,1,0)$ and $G(x;\, 1/2,1/2)$ in Fig.~\ref{Fig:slab}(b).  For 
a case when only the $1/2$-qv is stable inside the sample.
The energy difference between the two types of vortices at the center
is estimated to be much larger
than $k_BT_c$ at low enough temperatures (as large as
$10^{4}k_BT_c/\mu$m at $T\!\sim\!100$ mK). Hence, the entry of a
$1/2$-qv into the slab will be favored over that of a full qv.

However, the fact that many vortices will enter the slab at different
positions along its length complicates both
further theoretical modeling and experiments.  Thus, we also consider
cylindrical samples, which can only accommodate single vortices (of
either type) if the radius is small enough \cite{geim}.  For
example,  the fabrication of a sub-micron disk of ${\rm Sr_2 RuO_4}$, while challenging,
is within reach of available technologies such as focused ion beam (FIB) 
milling. The entry of a vortex or fluxoid in a cylinder or annulus
geometry can be detected as an  abrupt change in
the magnetization as $H$ is ramped up. As in the case of a thin slab,
a $1/2$-qv can be energetically favored for small enough
$\rho_{{\rm sp}}/\rho_{{\rm s}}$.

For simplicity,
we only compare the free energies for fluxoids trapped in a
thin, hollow cylinder. Although the fabrication of this geometry is
more challenging, it has an advantage over a
filled cylinder or disk of eliminating the core.
For a long, thin hollow cylinder of radius $R$ and thickness
$d \ll \lambda, R$,
the London approximation is nearly accurate~\cite{tony-victor}, and one can
obtain the Gibbs free energy per unit length for fluxoids  of
either type in the presence of an external field $H$ from Eq.~(\ref{EQ:gradientFree1}) via a Legendre transformation:
\begin{equation}
\frac{G(n_{{\rm s}},n_{{\rm sp}})}{\Phi_0^2/8\pi R^2}
\!=\!
\beta \left[\frac{1}{1\!+\!\beta}\left(n_{\rm s} \!-\! \frac{\Phi}{\Phi_0}\right)^2 \!+\! \frac{\rho_{{\rm sp}}}{\rho_{\rm s}} n_{{\rm sp}}^2
\right]\!-\!  \left(\frac{\Phi}{\Phi_0}\right)^2 \ .
\label{eq:G-cyl}
\end{equation}
We defined $\beta\!\equiv\! dR/2\lambda^2$\cite{geo_factor}, and
$\Phi\!\equiv\!\pi R^2 H$, the externally controlled flux through the 
cylinder. 
In the ground state, the values of $n_{{\rm s}}$ and $n_{{\rm sp}}$ minimize G for a given
value of $\Phi$. 
The $n_{{\rm sp}}^2$ term in Eq.~\eqref{eq:G-cyl}, which is
the only difference from the $s$-wave case, that allows for the
$1/2$-qv possibility of $n_{{\rm s}}\!=\!n_{{\rm sp}}\!=\!1/2$.  At
$\Phi/\Phi_0 \!=\! 1/2$, $G(1/2,1/2)\!<\!G(1,0)\!=\!G(0,0)$ if $\rho_{{\rm
sp}}/\rho_{\rm s} \!<\! (1 \!+\! \beta)^{-1}$ and a $1/2$-qv will have lower
energy than a full qv.  The vorticity could be
detected by monitoring the magnetization, as it was done for bilayer Al rings, which effectively form a two-component
SC~\cite{hendrik}.

{\it Conclusion} -- 
We  considered the vortex energetics of $p_x+ip_y$ SC in the London limit using a GL formalism.
For an isolated $1/2$-qv,  we noted the importance of controlling a logarithmically
divergent energy due to the  unscreened spin current.
While this divergence is regulated in a pair of
$1/2$-qv's of opposite spin winding, $\rho_{{\rm sp}}/\rho_{\rm s}$ needs to be sufficiently small in order  to make a finite
separation in such a pair  (meta-)stable. Given the fragile nature
of $1/2$-qv's in bulk samples, we propose using mesoscopic
geometries to look for $1/2$-qv's.

While for sufficiently small $\rho_{{\rm sp}}/\rho_{{\rm s}}$
the pairs of $1/2$-qv's can be energetically
stable in bulk samples and a single $1/2$-qv can be stable in  mesoscopic samples, a number of real material features
can affect the existence or
complicate the observation of $1/2$-qv's in Sr$_2$RuO$_4$.
Such features include
locking of the ${\bf d}$-vector within the RuO$_2$ plane, proliferation of
microscopic domain walls,  significant core energy effects,  and boundary scattering.
Nonetheless,
given the exciting promise associated with these exotic excitations,
our analysis forms a point of departure in the quest of realizing $1/2$-qv's in Sr$_2$RuO$_4$ or other candidate
materials for exotic superconductivity, such as UPt$_3$~\cite{upt3}.

\noindent {\bf Acknowledgments} We are grateful to  A.J.\ Leggett, M.\ Stone, K.\ Moler, V.\ Vakaryuk, A.\ Fetter, H.-Y.\ Kee, C.\ Kallin, S.\ Kivelson, P.\ Oreto for their insights and many helpful discussions.  SC has been supported
by University of Illinois Applied Mathematics Fellowship and NSF Grant
No.\ DMR 06-03528, HB by DOE under Contract No.\ DE-AC02-76SF00515,
and EAK by the Stanford Institute for Theoretical Physics.

\end{document}